\documentclass[prl,twocolumn,amsmath,superscriptaddress, showpacs]{revtex4}

\usepackage{amssymb}
\usepackage{graphicx}

\begin{document}
\title{Fission of a multiphase membrane tube}
\author{J.-M. Allain}
\email{jean-marc.allain@lps.ens.fr}
\affiliation{Laboratoire de Physique Statistique, Ecole Normale
Sup{\'e}rieure, 24 rue Lhomond, 75231 Paris Cedex 05, France.}
\author{C. Storm}
\affiliation{Physicochimie Curie,
Institut Curie Section recherche, 26 rue d'Ulm 75248 Paris Cedex 05, France}
\author{A. Roux}
\affiliation{Physicochimie Curie,
Institut Curie Section recherche, 26 rue d'Ulm 75248 Paris Cedex 05, France}
\author{M. Ben Amar}
\affiliation{Laboratoire de Physique Statistique, Ecole Normale
Sup{\'e}rieure, 24 rue Lhomond, 75231 Paris Cedex 05, France.}
\author{J.-F. Joanny}
\affiliation{Physicochimie Curie,
Institut Curie Section recherche, 26 rue d'Ulm 75248 Paris Cedex 05, France}
\date{\today}

\begin{abstract}

A common mechanism for intracellular transport is the use of controlled deformations
of the membrane to create spherical or tubular buds. While the basic physical properties
of homogeneous membranes are relatively well-known, the
effects of inhomogeneities within membranes are very much an
active field of study. Membrane domains enriched in
certain lipids in particular are attracting much attention, and in this Letter we investigate
the effect of such domains on the shape and fate of membrane tubes. Recent
experiments have demonstrated that forced lipid phase separation
can trigger tube fission, and we demonstrate how this can be understood
purely from the difference in elastic constants between the
domains. Moreover, the proposed model predicts timescales for fission that
agree well with experimental findings.
\end{abstract}



\maketitle

        Internal organization is one of the most intriguing
aspects of the cell. Living cells have to actively maintain gradients of all
sorts. Compartmentalisation and trafficking aid it in doing so,
and both processes extensively use membranes. Not
only are the various organelles in eukaryotic cells surrounded
by membranes, but the basic intermediates in the intracellular
transport pathways as well are membrane structures such as tubes and
vesicles \cite{Cell}. The generation and properties of these structures have been
extensively studied, and much is already known about their
biology, biochemistry \cite{Rustom04} and their
biophysics \cite{Seifert97, Roux02}. The emerging view is that the
shape of the bilayer membrane {\em in vivo} is controlled not only
by embedded and associated proteins \cite{Sciaky97} but also to a large extent by the
mechanical properties of the bilayer itself \cite{Upadhyaya04, Seifert97}. For
tubular structures in particular, mechanical effects play a major
role: recent biomimetic experiments \cite{Roux02} have shown that kinesin motors walking on
microtubules can exert pulling forces on the membrane and prompt the formation
of membrane tubes that resemble tubules identified in living cells.

The existence of small membrane domains with a lipid composition
that is markedly different from that of the rest of the membrane
(sometimes referred to as "rafts" although considerable debate remains
as to their precise interpretation) appears to be another
key element of intracellular vesicular traffic \cite{Huttner01}, and also seems
to be implicated in a multitude of cellular processes \cite{Simons97}.
The heterogeneity in membrane composition can be attributed to a phase
transition leading to a local segregation between the various lipids 
constituting the membrane \cite{Veatch03a}. Sphingolipid domains in particular
have been shown to be more structured than a classical liquid membrane
due to specific interactions between their constituents \cite{Cell}.
Under appropriate conditions they tend to aggregate into so-called
{\em liquid-ordered} domains which are mechanically
stiffer than the rest of the bilayer. Recently,
an experimental model system of vesicles including ``raft-like domains'' has
been developed \cite{Dietrich01}; it  provides an elegant and efficient tool to study their
properties in a more controlled way than {\em in vivo}. This procedure allows for
systematic studies of the effects of membrane composition
\cite{Wang00}, temperature changes \cite{Veatch03a} and protein absorption on the domain
\cite{Staneva04}.

The physics of membrane tube formation from homogeneous vesicles has been studied both theoretically
\cite{Powers02, Bukman96} and experimentally \cite{Heinrich99}.
\begin{figure}[t!]
        \includegraphics[width=0.4\textwidth]{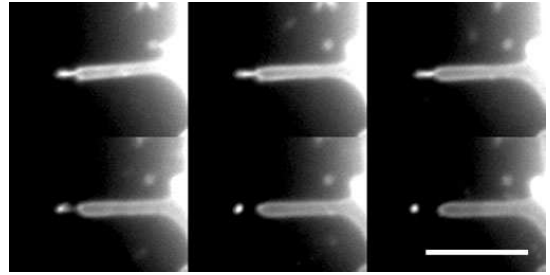}
        \caption{Breakage of a heterogeneous membrane
tube \cite{RouxPerso}. The brighter (and thinner) section at the tip on the left is a liquid-disordered DOPC domain. Fission events occur at the sites of
formation of small domains resulting from phase separation. The time between two consecutive
pictures is one second. Scale bar, $10\,\mu\text{m}$.
\label{Manip}}
\end{figure}
Recent experiments involving one of us \cite{RouxPerso} study the interplay between lipid
domains and the behavior of tubes, by pulling tubes from 
model membranes. Fig.~\ref{Manip} illustrates one of the suprising conclusions
of these experiments - a sequence of snapshots taken at regular
intervals (one second between two pictures) show an initially homogeneous tube that first undergoes
phase separation (triggered experimentally by photoinduced oxidization of cholesterol),
and, after about one second, ruptures precisely
at the phase boundary and disconnects. The two lipid phases are easily distinguished, once
separation has occurred, by the use of a fluorescent marker that
preferentially sits in the liquid-disordered domains. Furthermore,
the same experiments show that fission events such as these happen
{\em only} in
the phase separated tubes - tubes in which the lipids are mixed
are essentially stable indefinitely.

{\em Statement of problem and summary.}
In this Letter, we address the dramatic loss of stability following phase separation
from a mechanical point of view. We extend the theoretical models developed for
homogeneous tubes \cite{Bukman96} to study the junction
between two distinct phases, each of which far away from the
junction has a tubular shape.  Experiments suggest that phase
separation occurs on a much faster timescale than fission, and that the nucleation of the
two phases leads to the formation of cylindrical domains between a more rigid and
a less rigid phase. We therefore choose not to model the dynamic of the phase separation
process \cite{Chen97}. The tube radii and the junction length are generally small
compared to the length of each phase domain. In order to minimize the interfacial energy
between adjacent domains, the interface
rapidly becomes a circle
perpendicular to the tube direction. The coarsening stage of the
phase separation process proceeds very
slowly to eventually form two homogeneous phases in equilibrium,
but this slow relaxation is always preempted by tube
fission.

We assume here that the tube
and junctions are axi-symmetric with respect to the direction
along which the tube is pulled (the $z$ axis). We consider one junction 
between two semi-infinite tubes each consisting
exclusively of one of the phases.
Finite-size effects associated with the limited size of individual
domains, while possibly relevant, fall outside the scope of the present
paper. The small radius of the tubes (about $40$ nm) does not allow a quantitative determination
of the shape of the junctions \cite{RouxPerso}, and for this reason we restrict ourselves to a
minimal model which emphasizes the roles of the most relevant
physical parameters. We show that tube fission
can be driven either by the line tension or by the jump of the
elastic coefficients at the interface between the
two phases, and we compare the two processes that undoubtedly both contribute in the experiments.

\begin{figure}
        \includegraphics[width=0.4\textwidth]{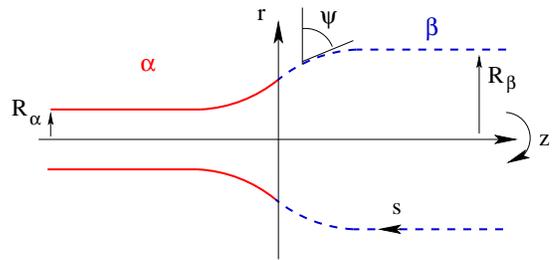}
        \caption{Schematic representation of the junction.\label{fig1}}
\end{figure}

{\em Model.} We use an elastic membrane free energy as introduced by Canham and
Helfrich \cite{Helfrich73}, and numerically determine
equilibrium junction shapes. Fig.~\ref{fig1} gives a schematic
representation of the tube and the coordinate system used in the following.
Our axisymmetric surface is parametrised by the arc length $s$ along the
contour and described by the local tube radius $r(s)$ and the
angle $\psi(s)$. They are related by the geometric relations $\dot
r=\cos\psi$ (dots denote derivatives with respect to $s$).
The interface is located at $z = s = 0$.

The free energy of the system is obtained by extending the description of
tubular membranes \cite{Bukman96} to the specific case of
a biphasic tube \cite{Julicher96a, Allain04} as follows:
\begin{eqnarray}
\label{Energie2}
{\cal F}\!=\! \!\!\sum_{i=\alpha ,\beta}\!
{\int _{\Omega_i} \! \left[ 2 {\kappa _i}{\sf H}^2 \!+\! \kappa
_{\text G}^{(i)}{\sf K}\!+ \sigma _i
\right] \!{\rm d}S}\!+\!\!\oint_{\partial \Omega}\!\!\!\tau\,{\rm d}\ell\!-\!\!\int\!\! f\,{\rm d}z.
\end{eqnarray}
The two phases are
denoted by $\alpha$ and $\beta$, and for each phase $i$
the free energy is integrated over its membrane area $\Omega_i$.
The $\kappa_i$ and $\kappa_{\text G}^{(i)}$ are the bending- and Gaussian rigidities of the
respective phases. This free energy includes the bending
energy to lowest order in the principal curvatures, where ${\sf H}$ is the mean curvature
and ${\sf K}$ the Gaussian curvature. The two layers of the membrane are
assumed to be symmetric - both phases contain cholesterol molecules which have
a high flip-flop rate. Any stress due to area differences
between the leaflets or to an asymmetry of the layers is thus quickly relaxed.
Finally, Lagrange multipliers $\sigma_i$ are introduced to ensure a constant surface in each phase. 
These $\sigma_i$ are interpreted as surface tensions. We take our tube to be infinite, and assume 
the presence of a lipid reservoir. In the experiments, such a reservoir is provided by the large 
mother vesicles from which the tubes are drawn. Provided the area per lipid remains constant during 
the process this implies a constant surface tension in each of the phases.

\begin{figure*}[ht]
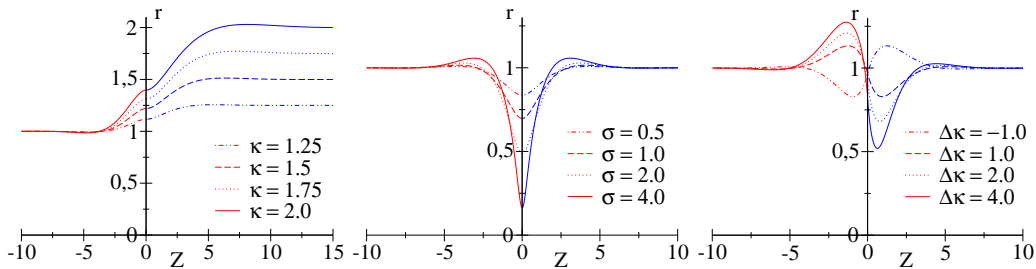

        \includegraphics[width=0.25\textwidth]{fig_kappa_3.eps}
        \includegraphics[width=0.25\textwidth]{fig_sigma_3.eps}
        \includegraphics[width=0.25\textwidth]{fig_gauss_3.eps}
        \caption{ Numerical shapes of the junction for various line
tensions and differences of elastic rigidities in dimensionless units.
The length scale is the radius of phase $\alpha$ ($R_\alpha = 1$), the energy scale is the bending
rigidity of phase $\alpha$ so that $\kappa_{\alpha} = 1$.
{\em (a)}: shapes for various ratios of bending rigidities. The line tension vanishes ($\tau = 0$)
and the Gaussian rigidities are equal ($\Delta \kappa_{\text G} = 0$). The values of
$\kappa _\beta /\kappa _\alpha$ are $1.25$, $1.5$, $1.75$, and $2.0$
{\em (b)}: shapes for various line tensions. The elastic rigidities are equal:
$\Delta \kappa_{\text G} = 0$ and $\kappa _\alpha = \kappa _\beta$. The values of the line tension
are $\tau = 0.5$, $1.0$, $1.5$ and $2.0$.
{\em (c)}: shapes for various differences in Gaussian rigidity. The line tension vanishes
($\tau = 0$) and the bending rigidities are equal ($\kappa _\alpha = \kappa _\beta$).
The values of the difference in Gaussian rigidity are
$\Delta \kappa_{\text G} = \kappa _G^{\beta} - \kappa _G^{\alpha} = -1.0$, $1.0$, $2.0$ and $4.0$.
\label{shapes}}
\end{figure*}

The interface between the two phases is described by
a jump in the values of the bending rigidities
$\kappa _i$, $\kappa _G^{(i)}$ and in the surface
tension $\sigma _i$, and by a positive line tension $\tau$ at the
interface $\partial \Omega$. The last term in the free energy is the work performed by the external
 force $f$ needed to pull the tube. We neglect the small effect of pressure \cite{Powers02}.

The variational derivation of the shape equations of the surface has been detailed elsewhere
\cite{Helfrich89}, and yields
\begin{eqnarray}
\label{EL1}
        \mathop \psi \limits^ {...} &=& -\frac{ {\mathop \psi
\limits^ {.}} { }^3}{2} - \frac{2 \cos \psi}{r}\mathop \psi \limits^
{..}
+ \frac{3\sin \psi}{2r} {\mathop \psi \limits^ {.}} { }^2  +
\frac{3\cos ^2\psi -1}{2r^2}\mathop \psi \limits^ {.} \nonumber \\
        && - \frac{\cos ^2 \psi +1}{2r^3}\sin \psi + \frac{\sigma}{\kappa}
\mathop \psi \limits^ {.}+\frac{\sigma}{\kappa} \frac{\sin \psi}{r}
\end{eqnarray}
Far away from the junction,
we recover homogeneous cylindrical tubes with $\psi = \pi /2$ and 
$R_i=(\kappa_i/2\sigma_i)^{1/2}$. Mechanical
equilibrium implies that the forces at both extremities are equal and that
$f=2\pi (2\sigma_i\kappa_i)^{1/2}$, which imposes
that $\sigma _\alpha / \sigma _\beta = \kappa _\beta / \kappa _\alpha$: The surface tension jumps discontinuously across the interface.

The mismatch between constants such
as the bending rigidities appears only in the boundary conditions and strongly
affects the interface shape. At the interface ($s = 0$),
four boundary conditions must be satisfied. Two conditions
are the continuity of the radius $r(s)$ and the angle $\psi(s)$ \cite{Julicher96a};
two additional conditions stem from the variational
procedure and relate the first and the second derivatives of the
angle $\psi$ on each side of the interface to the values
of $r$, $\psi$, $\kappa _\alpha$, $\kappa _\beta$, $\Delta
\kappa_{\text G} = \kappa _{\text G}^{\beta}-\kappa_{\text G}^{\alpha}$ and $\tau$.

{\em Results.} Fig.~\ref{shapes} illustrates the different effects that line tension and
differences in elastic rigidities individually have on the two-phase tube.
The first possible discontinuity at the junction is a jump in bending
rigidities (Fig. \ref{shapes}{\em (a)}). The ratio of the bending rigidities in the two phases
$\kappa=\kappa_{\beta}/\kappa_{\alpha}$ fixes the ratio of the radii away from the junction
and of the surface tensions in the two phases. In the absence of both line tension and jump in
Gaussian rigidity, the radius decreases smoothly from the values of the more rigid phase
to the value in the less rigid phase, but with a remarkable
structural feature - a small plateau ({\em i.e.} a membrane region with a
horizontal tangent) occurs around the junction. This plateau is also given by an analytical
linear calculation \cite{Futur}.

When line tension dominates (Fig.~\ref{shapes}{\em (b)}), the radius at the interface
decreases with increasing line tension. It vanishes for a huge line tension. Note that our 
description breaks down at scales comparable
to the bilayer width. Despite the fact that the radius goes to zero the mean
curvature remains finite; in the highly pinched limit a saddle point develops
at the neck which keeps the total curvature energy finite.

When the discontinuity in Gaussian rigidities dominates (Fig.~\ref{shapes}{\em (c)}),
numerical evidence suggests that the neck radius does not decrease all the way down to zero.
Moreover, stability arguments given below impose a bound on the maximum absolute value
of $\Delta \kappa _G$. However, the presence of the neck
favors the breaking process. In this case, fission does not occur exactly at the interface
but at the neck. One thus expects to find, after fission, a small patch of one phase still
attached to the other phase. Since details at the length scale of
the neck itself cannot be resolved experimentally, this effect
might be relevant to determine the dominant fission mechanism.

{\em Discussion.} For general experimental conditions, all three effects
are superimposed at the junction. A quantitative analysis of the shape in order
to extract the various parameters is then difficult, especially as
little to nothing is experimentally known about the precise shape
of the junction. Typical values of the bending rigidity of liquid
bilayers are around $25\,k_B T$, and the rigidity of the liquid
ordered phases can be up to several times higher. Recently, the bending modulus of a
heterogeneous vesicle has been obtained by
comparing the experimental shape to numerical solutions of the shape
equations \cite{Baumgart03}. The Gaussian rigidity $\kappa _{\text G}^{(i)}$ is notoriously difficult
to measure experimentally, but a recent study cites values of $\kappa _G^{(i)} = -0.83\kappa _i$ \cite{Siegel04}. Stability arguments impose that $-2\kappa_i < \kappa _G^{(i)} <0$.

\begin{figure}
        \includegraphics[width=0.45\textwidth]{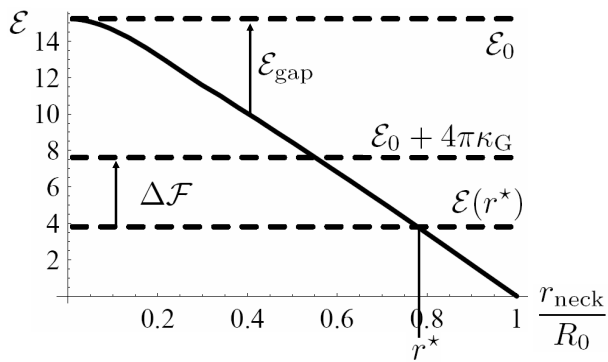}
        \caption{Schematic energetics of fission. The solid curve plots the
free energy of a tube pinched by line tension as a function of the
dimensionless neck radius. Every equilibrium radius $r^\star$ has
a corresponding energy ${\cal E}(r^\star)={\cal E}_{\rm bend}+{\cal E}_\tau$ which defines in turn
an energy barrier for fission ${\cal E}_\text{gap}$ and a free
energy gain upon fission $\Delta {\cal F}$. For clarity,
this figure assumes identical elastic rigidities on both sides.
\label{barriere}}
\end{figure}
The equilibrium free energy of the tube can be calculated from
Eq.(\ref{Energie2}) and allows a discussion of the stability
of the tube and of its fission. We show in Fig.~\ref{barriere}
the free energy of a tube as a function of the dimensionless radius at the neck
$r_{\text{neck}}/R_0$ in the specific case of $\kappa _\alpha = \kappa _\beta$ and
$\kappa _G^{\alpha} = \kappa _G^{\beta}$. This energy is maximal for a vanishing
radius: at this point, the membrane is maximally bent. Fission of the tube by pinching requires one to cross this energy barrier. The free energy of the ruptured tube is also shown on the figure. It is lower than the top of the barrier by the contribution of the Gaussian curvature due to the change in
topology upon rupture, which equals $4\pi \kappa_{\text G}$. Notice that the bending energy does {\em not} change upon rupture: at vanishing radii, the neck is a saddle point with vanishing mean curvature \cite{Futur}. The ruptured tube corresponds to a transient shape since in the absence of an applied force, the tubes retract to form two spheres. In the absence of line tension  the tube is uniform $r(z) =R_0$ and its energy
is zero, and a homogenous tube is thus thermodynamically stable
only if the free energy of the ruptured tube is positive. Numerically, we have
determined this stability limit as $\kappa_{\text G}>-1.29\kappa$.

The values of the parameters then fix the value of  $r_{\text{neck}}/R_0$. We have also evaluated the energy barrier against fission by pinching from this macroscopic model. Note, however, that this is only a lower bound to the real energy barrier, as it ignores effects at the molecular lengthscale which certainly is attained when the neck becomes very thin.
To compare our results to the experiments, we have computed the various energies at the following (measured or realistic) parameter values. With bending rigidities $\kappa_\alpha=40\,k_B T = 1.6\cdot 10^{-19}
\,\text{J}, \kappa_\beta=70\,k_B T=2.9\cdot 10^{-19}\,\text{J}$,
Gaussian rigidities $\kappa _G^{\alpha} = -33.2\,k_B T=-1.38\cdot 10^{-19}\,\text{J},
\kappa_G^{\beta}=-58\,k_B T=-2.3\cdot 10^{-19}\,\text{J}$, surface tensions
$1\cdot 10^{-6}\,$N/m in phase $\alpha$ and $5.7\cdot 10^{-7}\,$N/m in phase $\beta$,
and a line tension $7 \cdot 10^{-12}\,$N, we have determined the
height of the energy barrier to be ${\cal E}_{\text{gap}}=7.8\,k_B T$.
If we assume that fission is a thermally activated process \cite{Pomeau92},
the average time until fission $t_b$ occurs is $t_b = t_0 \exp{{\cal E}_{\text{gap}}/k_B T}$.
Using a hydrodynamic argument, we estimate the basic time scale as
$t_0=\eta R_\alpha^3/\kappa_\alpha$, where $\eta$ is the viscosity of water. For the parameter
values cited above this yields a timescale $t_0\approx 1.44\cdot10^{-4}\,$s. We thus expect
the experimental time until fission to be approximately $350\,$ms. This is in good agreement
with the experimentally observed typical time for fission, which is of order $1$ s.

{\em Conclusion.} We have studied the behavior of a multiphase membrane tube using thermodynamic
arguments. The shape of the junction between two domains depends on three quantities: the
line tension of the interface and the jumps in the two elastic constants. While experimental
precision is not yet at a level where these results can be compared directly to our calculated tube
shapes, we have also considered the breaking time of a two-phase tube. Our modeling, based on an energetic
approach, predicts a strong dependence of the fission dynamics on the elastic properties of the
phases and yields results that are in good agreement with the experimental data.

{\em Acknowledgments.} We would like to thank Patricia
Bassereau, Bruno Goud and Jacques Prost for stimulating discussion
and suggestions. C.\,S. acknowledges support from the European
PHYNECS research network.

\bibliographystyle{unsort}

\end{document}